\documentclass[aps,prresearch,twocolumn,groupedaddreApplyingss,nofootinbib,10pt,longbibliography]{revtex4-1}
\usepackage{CJK}

\usepackage{hyperref}
\usepackage{xcolor}
\usepackage{braket}
\usepackage{slashed}

\usepackage{mathtools}
\usepackage{multirow}
\usepackage{nicefrac}
\usepackage{amssymb}
\usepackage{mathrsfs}
\usepackage{IEEEtrantools}
\usepackage{tabularx}

\definecolor{BrickRed}{cmyk}{0, .89, .94, .28}
\definecolor{MyDarkBlue}{rgb}{0,0.08,0.45}

\newcommand{\be}{\begin{equation}}
\newcommand{\ee}{\end{equation}}
\newcommand{\tens}[1]{\stackrel{\leftrightarrow}{#1}}

\begin{document}

\title{Multiscale modeling of resistive switching in gold nanogranular films}

\author{Miquel L\'opez-Su\'arez}
\email{mlopez@dsf.unica.it}
\affiliation{Dipartimento di Fisica, Università degli Studi di Cagliari, Cittadella Universitaria, I-09042 Monserrato, Cagliari, Italy}
\author{Claudio Melis}
\email{claudio.melis@dsf.unica.it}
\affiliation{Dipartimento di Fisica, Università degli Studi di Cagliari, Cittadella Universitaria, I-09042 Monserrato, Cagliari, Italy}
\author{Luciano Colombo}
\email{luciano.colombo@dsf.unica.it}
\affiliation{Dipartimento di Fisica, Università degli Studi di Cagliari, Cittadella Universitaria, I-09042 Monserrato, Cagliari, Italy}
\author{Walter Tarantino}
\email{wtarantino@dsf.unica.it}
\affiliation{Dipartimento di Fisica, Università degli Studi di Cagliari, Cittadella Universitaria, I-09042 Monserrato, Cagliari, Italy}

\date{\today}

\begin{abstract}
Metallic nanogranular films display a complex dynamical response to a constant bias, showing up as atypical resistive switching mechanism which could be used to create electrical components for neuromorphic applications. To model such a phenomenon we use a multiscale approach blending together an ab initio treatment of the electric current at the nanoscale, a molecular dynamical approach dictating structural rearrangements, and a finite-element solution of the heat equation for heat propagation in the sample. 
We also consider structural changes due to electromigration which are modelled 
on the basis of experimental observations on similar systems.
Within such an approach, we manage to describe some distinctive features of the resistive switching occurring in nanogranular film and provide a physical interpretation at the microscopic level.
\end{abstract}

\maketitle


\section{Introduction}
\label{sec:intro}

The capability of engineering systems down to the atomic level
reached in the last decades has open the door 
to the possibility of an unprecedented tailoring of material properties.
Newly synthesized nanostructured materials can in fact also present 
features bearing the potential of unforeseen technological applications.
This is the case, for instance, of cluster-assembled gold thin films,
which present a rich dynamical non-Ohmic response to an external bias
that could be exploited, it has been suggested \cite{mirigliano2019,mirigliano2020a,mirigliano2021b}, 
in the fabrication of unconventional computing hardware components.
Even under a constant bias, the electrical resistance of such systems
can suddenly change over macroscopic scales giving rise 
to what is usually called resistive switching events. 
The microscopic mechanisms behind them are still far from being identified 
and fully characterized.
Partially, this is due to a lack of specialized theoretical tools
that allow to carry out a quantitative analysis of the phenomenon.
In two previous works, the first steps towards the definition of such tools
have been moved \cite{tarantino2020,lopez2021}, and a further one is presented here.

More specifically, we propose to model the electrical response of such systems
with a multiscale approach.
At the macroscopic scale a set of classical equations, 
Fourier heat equation and Kirchhoff's circuit laws coupled via Joule heating,
are numerically solved on a uniform grid; 
they allow to estimate the film electrical resistance and temperature.
At the microscopic scale, molecular dynamics simulations
and an \textit{ab initio} tool previously developed \cite{lopez2021}
are used to estimate the value of electrical conductivity
that enters into the classical equations.
This allows us to treat such a quantity in its general form
as a temperature-dependent tensor field, 
reflecting the anisotropic and non-homogeneous nature of the sample at small scales.
In fact, we shall argue that for such systems 
electrical conductivity sensibly depends on the current as well
and propose a model based on observations of electromigration effects in gold nanowires.

In Section \ref{sec:macro} we will present our theoretical device in detail;
after discussing some motivations, we shall introduce the macroscopic equations
that dictate the evolution of the observables of interest;
then we will discuss how theory and experiments can help us in estimating
the highly nontrivial values that a temperature- and current-dependent
inhomogeneous electrical conductivity can assume in those systems.
In Section \ref{sec:results} we will present the results of 
a simulation of the entire procedure.
Specific features of the experiments will emerge from the few ingredients
of the model, allowing us to suggest an explanation about their microscopic nature.
Conclusions and outlook will be presented at the end.

\section{Multiscale modeling}
\label{sec:macro}
\subsection{Experiments and modeling, an overview}
\label{subsec:motivs}
Purely metallic clusters of a few nanometers of diameter 
can be produced in gas phase and gently deposited 
on a substrate to form, layer after layer, a ``nanogranular film'' (or \textit{ng-film}),
i.e. a large ($\sim$mm$\times$mm),
uniform agglomerate of clusters that retain their individuality to a high degree.
By growing a ng-film between two electrodes, 
one can probe its response to an external electrical bias \cite{barborini2010}.
Besides the usual insulating/metal transition occurring around the percolation threshold (a few nanometers),
gold ng-films have been reported to present a complex non-Ohmic response even 
when the film has grown far from that threshold (tens of nanometers).
In the simplest situation of a constant bias, the response of the film is highly dynamical.
More specifically \cite{mirigliano2019,miriglianothesis},
upon application of a sufficiently high bias, the resistance of a ng-film, 
which starts with a low, constant value,
suddenly increases to a much higher value (two orders of magnitude);
after this initial phase, called of ``sample activation'', 
resistive switching events show up, appearing 
in the form of sudden ``jumps'' of the sample resistance;
while jumps occur over macroscopic scales (seconds), 
the jump itself is characterized by a much shorter time scale (fraction of second);
they can occur in either directions: to higher as well as to lower values;
recurrent resistance values can be observed over long observational periods;
the higher the voltage of the external bias the more frequent the jumps are;
sample activation is permanent: after the bias is switched off,
an activated sample will show right away resistive switching events 
the next time a bias is switched on.

What causes such a form of resistive switching is far from clear.
Different mechanisms at different scales may occur;
at the atomic scale, local charging/discharging phenomena may happen;
at the cluster scale, the structure of single clusters may vary 
(defect migration, local melting or crystallization,...);
at a larger scale, the configuration of groups of clusters may change over time 
(coalescence, aggregation, sliding, densification...).
A cyclic mechanism of electric current bridges forming/breaking at the cluster interfaces 
has been suggested \cite{mirigliano2020a} but not yet supported with sufficient
experimental evidence or a convincing quantitative theoretical analysis.

From the theoretical side, indeed, the phenomenon poses some challenges,
specifically for its dynamical character.
In a static situation, the resistance of a ng-film can be estimated using a variety of approaches.
If the ng-film is simply regarded as a thin film, one can use Matthiessen’s rule and 
conceptualize its resistivity as that of bulk Au plus the sum of different contributions; 
for most of them (grain boundaries, surface effects,...) 
there are well-developed models are routinely used for polycrystalline thin films \cite{mirigliano2021a}, 
while some other (interstitial voids, random grain orientation,...) may require an \textit{ad hoc} treatment \cite{tarantino2020}.
Regarded as a granular system, the static resistance of ng-films can be estimated
adapting analytical and numerical tools from percolation theory \cite{kirkpatrick1973,stauffer1994,sahimi1994},
or, for regimes far from the percolation threshold, those developed for ordered grain arrays
based on Green's function formalism \cite{beloborodov2007}, which well capture inter-grain quantum effects.
If regarded as a inhomogeneous medium, its resistance can be estimated using effective medium approximations \cite{landauer1978,stroud1998}
which can also accurately describe quantum effects \cite{grimaldi2014}. In Ref. \onlinecite{lopez2021} we presented
a procedure based on an atomistic modeling of the film via 
molecular dynamics and a characterization of the electric transport
via ab initio techniques which, although computationally expensive,
provided us with a flexible tool to estimate the resistance of ng-films.

Although all of the mentioned approaches can, in principle, 
be made time-dependent, not all of them are flexible enough be model any possible 
microscopic mechanism. 
Consider, for instance, the approach we presented in Ref. \onlinecite{tarantino2020}.
To model the film in a dynamical situation we used
a time-dependent effective medium approximation
with which we were able to prove that local resistance changes
can lead to global ones if they are triggered by the temperature;
as the sample heats up due to Joule effect,
resistive switching events at the sample scale
appear in correspondence of some sort
of phase transition, during which certain local structural arrangements 
become more likely then others.
The model, however, relies on certain assumptions that prevent us to study
other possible situations of interest. For instance,
the underlying assumption of homogeneity of charge and heat diffusion
prevents us to explore what happens if the current tends to circulate only 
on a few preferred percolation paths. 
Or, the independence of local resistance variations that are randomly decided
prevents us to model the complex interplay between current and temperature 
at the cluster scale required by the above mentioned 
cyclic mechanism put forth in Ref. \onlinecite{mirigliano2020a}.

In the present work we introduce a new approach designed to
overcome these limitations.
Using molecular dynamics simulations to model ng-films
freed us from the constrains of dealing with clusters 
as fundamental brick of the modeling.
Our system was indeed not just a network of pristine clusters,
but a rich structure of randomly oriented grains, 
amorphous interfaces, interstitial voids, etc.
This is a natural framework for describing local melting,
crystallization and all sort of unbiased structural changes 
can arise at the cluster scale.
However, molecular dynamics modeling comes with its own limits
and one cannot directly introduce an electric current 
and see its effects on the atomic structure,
simply for the lack of electrons in those simulations.
Even bypassing the current and simply looking at its effects
on an atomic structure due to Joule heating 
is a procedure that presents its conceptual difficulties:
for structures as such where charge transport is often 
in its ballistic regime \cite{lopez2021},
it might be hard to establish which atoms are heated up by Joule effect (cf Ref. \onlinecite{bringuier2017} and references therein);
and even if the Joule heat was appropriately distributed,
its propagation through neighboring atoms,
in great part due to electronic degree of freedom,
would not be correctly captured
by molecular dynamics simulations, 
in which electrons are not included.
A more sensible way to deal with Joule heating in atomistic simulations
is to couple them to a resolution of an appropriate heat diffusion model
\cite{padgett2005,crill2010,donati2020}.
And this is the strategy we present here.

\subsection{The non-homogeneous, anisotropic thermistor problem}
\label{sec:thermistor}

When a bias is applied to a ng-film, such system can respond in a non-linear way.
This is probably the result of structural changes happening 
at the cluster level and triggered by local variations
of temperature and current.
Key observables are therefore the temperature and vector current field,
$T({\bf r},t)$ and $\vec{I}({\bf r},t)$, respectively.
At a macroscpic level they can be described via what is sometimes called 
``the thermistor problem'' \cite{antontsev1994},
namely Kirchhoff's circuit law and the Fourier heat equation with 
Joule heating as source term, 
complemented by appropriate initial and boundary conditions.
In a non-homogeneous, anisotropic situation these equations can be written as
\be\label{eq:fourier}
\rho c\frac{\partial T}{\partial t}=
    \vec{\nabla}\cdot\left(
    \tens{\kappa} \cdot \vec{\nabla}T\right)
    +\vec{I}\cdot\tensor{\sigma}\cdot\vec{I},
\ee
\be\label{eq:kirchhoff}
\vec{\nabla}\cdot\vec{I}=0
\ee
where $\rho$ is the mass density, $c$ the specific heat capacity,
$\tensor{\kappa}$ the thermal conductivity,
$\tensor{\sigma}$ the electrical conductivity,
and the notation $\vec{v}\:\cdot\!\! \tens{M}\!\! \cdot\: \vec{w}$ 
is used as a shorthand for $\sum_{a,b}v_a M_{ab}w_b$.
The source term in Eqn. \eqref{eq:fourier} is obtained combining $\vec{I}\cdot \vec{E}$ \cite{antontsev1994}
and $\vec{I}=\tensor{\sigma}\cdot\vec{E}$, $\vec{E}$ being the electric field.
Just like its homogeneous, isotropic counterpart,
Eqn. \eqref{eq:fourier}, which determines $T({\bf r},t)$, requires an initial value
and some boundary condition, which can be Dirichelet, 
Neumann,
or a mixture of the two; on the other hand,
Eqn. \eqref{eq:kirchhoff}, which determines $\vec{I}({\bf r},t)$,
only requires the boundary conditions that are dictated by the request
that the electric potential has the bias difference at the electrodes.
Details on the numerical resolution of such a set of equations is provided in the Appendix.

The complex microscopic structure of a ng-film enters, therefore,
via the quantities, $\rho$, $c$, $\tensor{\kappa}$, and $\tensor{\sigma}$.
By approaching the numerical resolution of the equation with a finite element method,
we discretize the space and partition it into a mesh.
Those observables are therefore directly related to the mass,
heat capacity, thermal and electrical conductance of each mesh cell.
To simplify the problem, we consider a mesh large enough for 
$\rho$ to be constant (i.e. space and time independent).
The existence of such a scale is justified by the fact that 
during the experiments no appreciable changes 
to the topography of ng-film samples has been reported:
even though, when a bias is applied, connections between clusters may form or deteriorate,
the density of clusters inside the film seems to remain constant.
Similarly, we also consider constant the specific heat capacity, which is primarily connected 
to the cell mass and not to the arrangement of clusters inside a cell.\footnote{%
While both $\rho$ and $c$ can be estimated using molecular dynamics simulations,
$\rho$ can also be estimated using Au bulk density and a plausible porosity 
(cf \cite{tarantino2020}),
while $c$ can be estimated using the Dulong–Petit law, 
again adjusted to account for the porosity of the medium.}
What can sensibly depend on the cluster arrangement inside a cell are the 
cell electrical and thermal conductance.
For cells containing a small number of clusters (say, a few units),
it is reasonable to expect that even a single junction, that for some reason breaks,
can lead to a sizeable change of electrical/thermal conductance in that direction. 
We therefore consider 
$\tensor{\sigma}=\tensor{\sigma}({\bf r},t)$ and 
$\tensor{\kappa}=\tensor{\kappa}({\bf r},t)$.
As we are dealing with a purely metallic system,
the thermal conductance of a cell can be estimated from the electrical conductance using the Wiedemann-Franz law, therefore we consider
\be\label{eq:wiedemannfranz}
\tens{\kappa}=L T \!\tens{\sigma},
\ee  
$T$ being the temperature field and $L$ Lorenz number.

It follows that the key quantity, connecting the macroscopic level of the model
to the lower one, is the electrical conductivity $\tensor{\sigma}({\bf r},t)$.
Such an observable indeed encodes whatever changes happens at a ``microscopic'' scale,
which in this context means ``at the scale of a single cell''.
Contrary to other common multiscale approaches,
we do not opt to calculate such an information at each timestep 
of the resolution of the macroscopic equation \eqref{eq:fourier},
but rather use theoretical tools and available experiments
to build, once for all, a map that connects $\tensor{\sigma}$ with $T$ and $\vec{I}$
and use that in the resolution of the closed set of equations 
\eqref{eq:fourier}, \eqref{eq:kirchhoff}, \eqref{eq:wiedemannfranz}.

\subsection{Temperature-induced conductivity variations}
\label{sec:temp}

As we go from the macroscopic level of the model to the microscopic one,
we consider the cell resistance, $R$, from which $\tensor{\sigma}({\bf r},t)$ 
is readily obtained.\footnote{%
The symbol $R$ should carry multiple indices:
one for the cell, one for the timestep and one for the direction 
on which the tensor is projected (in our case $i=x,y,z$), 
but we shall avoid such a complicated notation unless necessary.}
For the equations characterizing the macroscopic level to make sense,
each cell must includes (at least) a few gold clusters.
This means that, in estimating the cell resistance,
we can neglect the contribution of phonons, quantum tunneling, etc.
and focus only on structural changes at the atomic level which,
as argued in Ref. \onlinecite{tarantino2020,lopez2021}, 
are the main responsible for large resistance variations.
As explained in Ref. \onlinecite{lopez2021}, in doing this
the only relevant quantum effect to take into account is 
the complex interplay between ballistic and diffusive electronic transport regimes
between crystal and amorphous regions inside the system.
Molecular dynamics (MD) simulations and the \textit{ab initio} 
tool presented in Ref. \onlinecite{lopez2021}
(an atomically-resolved equivalent resistor network, or AR-ERN)
seem therefore adequate tools to estimate a cell resistance. 

\begin{figure}
\includegraphics[width=0.48\textwidth]{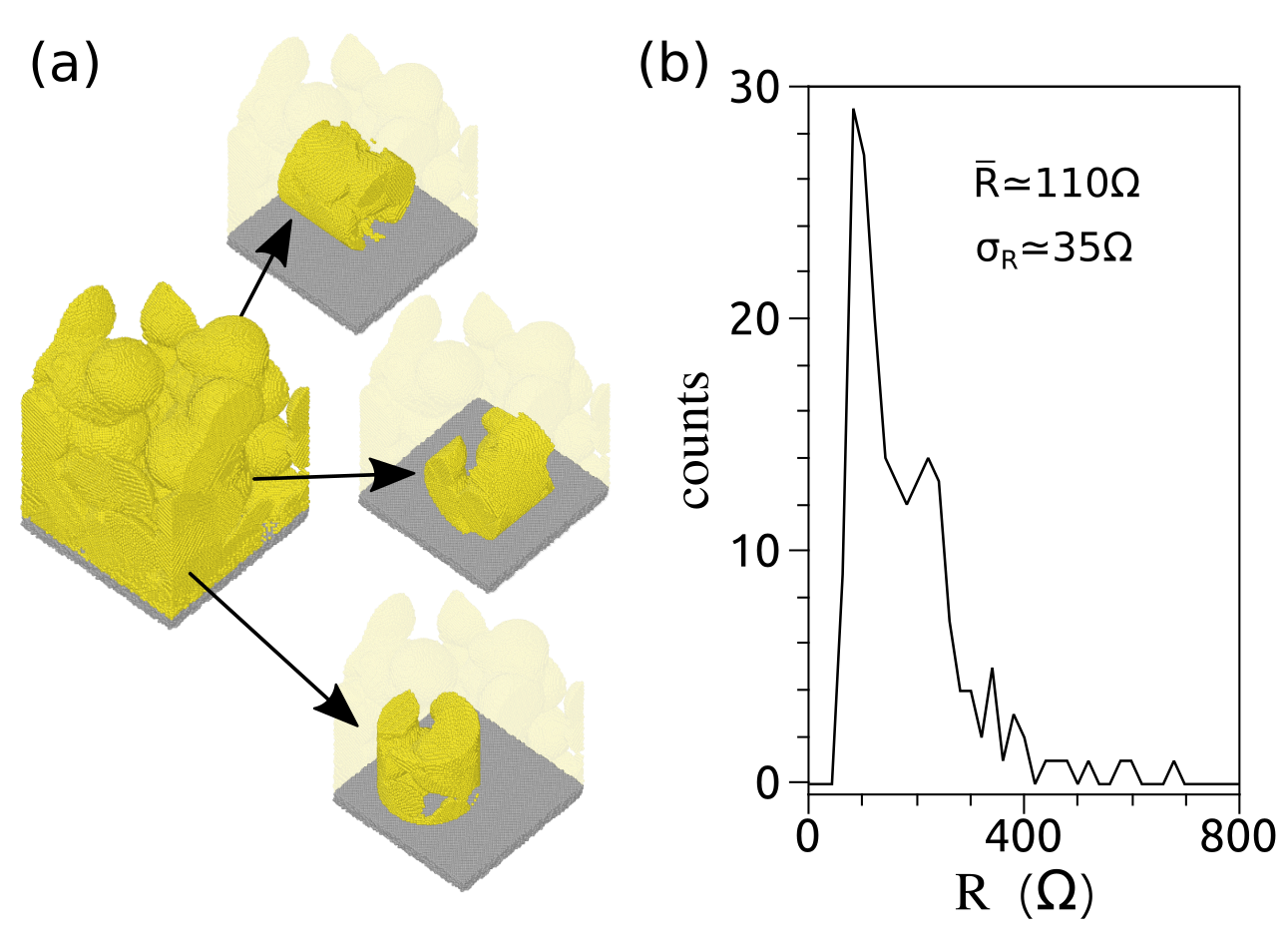}
\caption{%
To study as a cell resistance varies with the temperature 
we randomly carve samples out of a large scale MD simulation of a chunck of ng-film (left panel). 
As cubic samples have the tendency to deform when isolated from the surrounding system, 
we consider cylindrically shaped samples which present higher stability. We use periodic boundary conditions while cylinders are surrounded by vacuum so to effectively isolate periodic images of the cylinders. To represent cells of size 15nm
we consider cylinders of length 15nm and section diameter 15nm 
(right panel).
In calculating the resistance, electrodes are placed on the two opposite flat faces.
By taking many samples, one can estimate the random distribution of 
the cell resistance for a given ng-film (right panel).
}
\label{fig:samples}
\end{figure}

\begin{figure}
\includegraphics[width=0.48\textwidth]{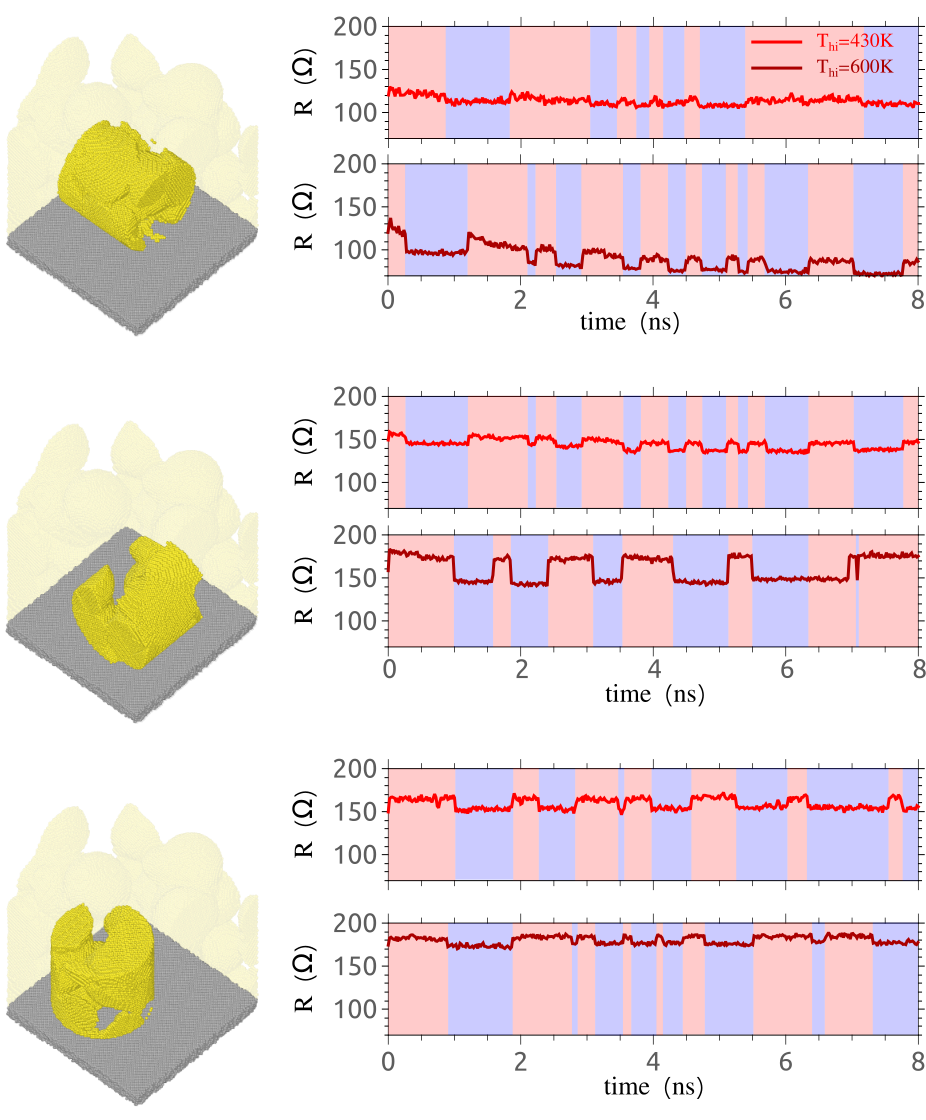}
\caption{%
Evolution of the resistance of three carved out samples
in connection with temperature variations.
For each sample, the resistance has been monitored
while changing the temperature. Blue shadowed regions correspond to the time intervals in which the temperature is 300K while red ones correspond to $T=430$ K (upper panels) and $T=600$ K (lower panels).
}
\label{fig:mdresults}
\end{figure}

\begin{figure}
\includegraphics[width=0.48\textwidth]{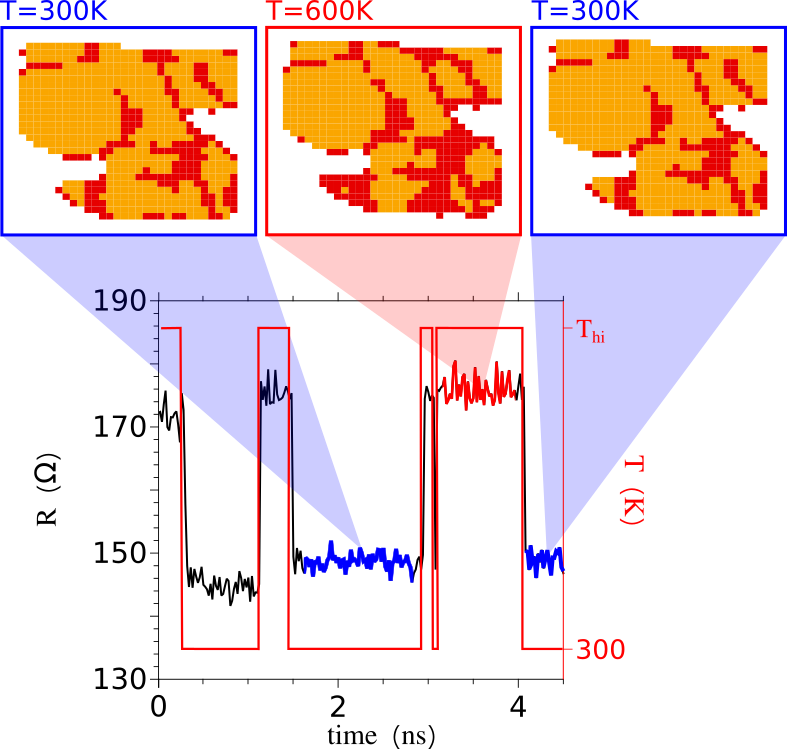}
\caption{%
Non-hysterical behaviour of the resistance as a function of the temperature.
As one can see from the highlighted regions,
the rise in temperature leads to an amorphization of a certain region,
while the temperature fall favours a re-crystallization of the same region.
The process seems reversible and justifies our assumption
that for each $T$ a single value of $R$ can be assigned to the system.
}
\label{fig:cycle}
\end{figure}

Next, starting from the large scale simulation of a chunck of a ng-film, 
as described in Ref. \onlinecite{lopez2021},
we carve out of the entire system a few samples
of the size of the cell of the macroscopic level ($\sim 15$ nm),
as shown in Fig. \ref{fig:samples}, left panel.
For the reasons mentioned in Section \ref{subsec:motivs}, we warm up the samples globally, 
rather then locally, and what we observe in general is a change of the structure
of the samples even when the highest temperature 
is considerably lower than gold melting point ($T_{melt}\sim 900\;K$). 
As one would expect, those changes depend on the specific function $T(t)$ one uses. 
However, when it comes to the electrical resistance of the sample,
we do observe that, after a brief period of adjustment,
if the sample is brought to a certain temperature $T=\tilde{T}$,
its resistance will roughly be $R=\tilde{R}$, no matter how $\tilde{T}$ is achieved.
This allows us to establish that, even though some hysteresis effects are present,
it is indeed possible to describe $R$ as function of $T$.
This is well exemplified by Fig. \ref{fig:mdresults}, 
in which we report the behavior of three samples
under a complex temperature evolution.
By randomly switching the sample temperature between room temperature (300 K)
and a certain higher temperature (430 or 600 K), 
all three samples tend to respond in the same way:
after a period during which the structure relaxes a bit,
to each temperature switching correspond a resistance switching
and a relation between specific values of $T$ and $R$ emerges,
allowing us to consider a function $R(T)$ indeed.
On our limited set of samples considered, the exact form of $R(T)$ 
sensibly depends on the sample, but we do recognize a trend that 
we encode in the following function:
\be\label{eq:RofT}
R(T) = R_{0}(1 + \vartheta(T-\tilde{T}) \Delta )
\ee
where $R$ is the resistance of a specific cell,
$T$ the cell temperature, $R_{0}$ the resistance of the cell at room temperature,
$\tilde{T}$ a certain threshold temperature (with $T_{room}<\tilde{T}<T_{melt}$) that depends on the cell, 
$\Delta $ a quantity that indicates a resistance variation,
also depending on the specific cell,
and $\vartheta$ the Heaviside function. In words,
Eqn. \eqref{eq:RofT} encodes a sudden change of the cell resistance
that might occur when the cell temperature is brought above a certain threshold.
Such a behavior seems to be linked to cycles of amorphization/re-crystallization 
happening in key regions as cluster interfaces, as shown in Fig. \ref{fig:cycle},
but further investigations would be required to make a general statement.

Based on our simulations, we give the rough estimates of 
$\tilde{T}\sim 400 \div 600 K$ and $\Delta \sim 0 \div 0.2$.
Providing more accurate ones would require the study of  $R(T)$
for many different cells under many different $T(t)$ profiles,
which, for its computational cost, is currently the scopes of the present work.
A much more accurate estimate of $R_{0}$, however, can be already given;
by considering a large set of samples carved out of the system of Ref. \onlinecite{lopez2021},
which is indeed at room temperature,
we recognise that the $R_{0}$ roughly follows a Poissonian 
distribution peaked around $110\; \Omega$,
as shown in Fig. \ref{fig:samples}, right panel.

Technical details about the MD simulations and the resistance estimates 
are provided in the Appendix.

\subsection{Current-induced conductivity variations}
\label{sec:curr}

As the dynamical observables of the macroscopic model are $T$ and $\vec{I}$,
we could speculate about a dependency of $\tensor{\sigma}$ %
not only on $T$ but on $\vec{I}$ as well.
In fact, we believe that in our case this is not only possible but actually necessary.
When a current high enough is sustained over time, the structure of the conductor 
may deteriorate as effect of electromigration (EM),
an effect that has been widely studied 
\cite{latz2012simulation,stahlmecke2006electromigration,knowlton1997simulation,schimschak2000electromigration,suo1994electromigration,sanchez1990morphology}
and reported also for poly-crystalline Au nanostructures
\cite{stahlmecke2007resistance,wu2007feedback,karim2009diameter}. 

A back-of-the-envelop calculation shows that such an effect is relevant for our case, too.
In Ref. \cite{karim2009diameter}, Au nanowires are reported to completely break 
after having transmitted high enough currents. According to the authors,
the highest value they can withstand before undergoing failure 
can be written as $I_{f}=6\times 10^3  A_c^{0.425}$ A,
where $A_c$ stands for the nanowire cross-section.
Such an expression has been confirmed for cross-sections ranging
from $5 \times 10^5$ nm$^2$ down to $8 \times 10^3$ nm$^2$.
In the case of the ng-films of Ref. \cite{mirigliano2019},
the cross-section is much higher (a typical ng-films has length and width equal to $1$ mm
and a thickness of a few tens of nm) and the intricate structure at the cluster scale
makes the section of a ng-film quite different from that of a nanowire 
(and more akin to a section of a random stack of nanowires).
Nonetheless we can use the same formula to get an idea of the order of magnitude 
of the failure current for those systems, which turns out to be $I_f=0.28$ A, 
a value that is not that far from the typical 
current in their experiments, $
I\sim 20$V$/100\Omega=0.2$ A.
We therefore argue that such an effect may be present and must be taken into account.

Because of the interplay between electrons and atoms,
it is hard to simulate the effects of electromigration on a generic structure
with standard computational tools,
since they are designed to treat the dynamics of either electrons 
or atoms.
Nonetheless, to our purposes, a simple characterization based on available experiments
is enough to give us some interesting effects in the macroscopic model.
More specifically, we consider the work of Ref. \onlinecite{wu2007feedback},
in which the degradation of Au junctions under the effects of different current regimes 
is systematically studied, and propose a simple way to reproduce 
their results within our macroscopic model.

When a Au junction is exposed to high biases for a long period, 
the material can develop some discontinuity in its structure that affects the flow of electrons.
Such an effect can be quantified by measuring the sample resistance
against the bias applied in a quasi-static experiment
during which the bias is slowly increased over time.
It is then possible to identify three regimes.
For low enough voltages, the sample acts in a Ohm-like manner,
as one would expect from a standard conductor in normal working conditions.
For moderate voltages, electromigration effects kick in
and the junction starts shrinking;
the beginning of this phase is marked by a sudden fall of the junction conductance,
in complete analogy with the activation of a ng-film.
For high enough voltages, the junction reaches a failure point
after which it no longer can sustain a current:
the junction is effectively broken.

An experiment as such can be, not only qualitatively, but also quantitatively reproduced 
in our modeling if an appropriate $R(I)$ is considered.
For start, we neglect temperature variations and consider only the circuit equation, 
Eq. \eqref{eq:kirchhoff},  with a potential difference at the electrodes 
that linearly increases in time. 
At each time step the current $I$ of each cell is calculated.
If such a value overcomes a certain (``activation'') 
threshold $I_a$, 
which is randomly assigned for each cell from a given range,
we say that the current shrinks the effective cross-section $A_c$ 
that the it experiences inside the cell.
Such a shrinking is quantified by a certain parameter $\beta$,
such that $A_c\to A_c':=\beta^{-1}A_c$.
According to Ref.  \cite{lopez2021modeling},
this implies an increase of the cell resistance $R\to R':=\beta R$.
At the next time step, the new values $R'$ and $A_c'$ 
will be considered, alongside some new values for the current
activation and failure, for each cell.
Note that we allow $A_c$ to be reduced down to $0.16$ nm$^2$
(corresponding to a minimal junction with area $a^2$, 
being $a$ the lattice constant of gold) and
we use $\beta=5$ to produce all the results in this work. 
If the cell current overcomes the failure threshold $I_f$,
which is also initially assigned to each cell at random from a certain range,
then the cell is considered ``failed'': 
a junction along the percolation path inside the cell breaks 
and no current is allowed to flow through that cell anymore.

\begin{figure}
\includegraphics[width=0.48\textwidth]{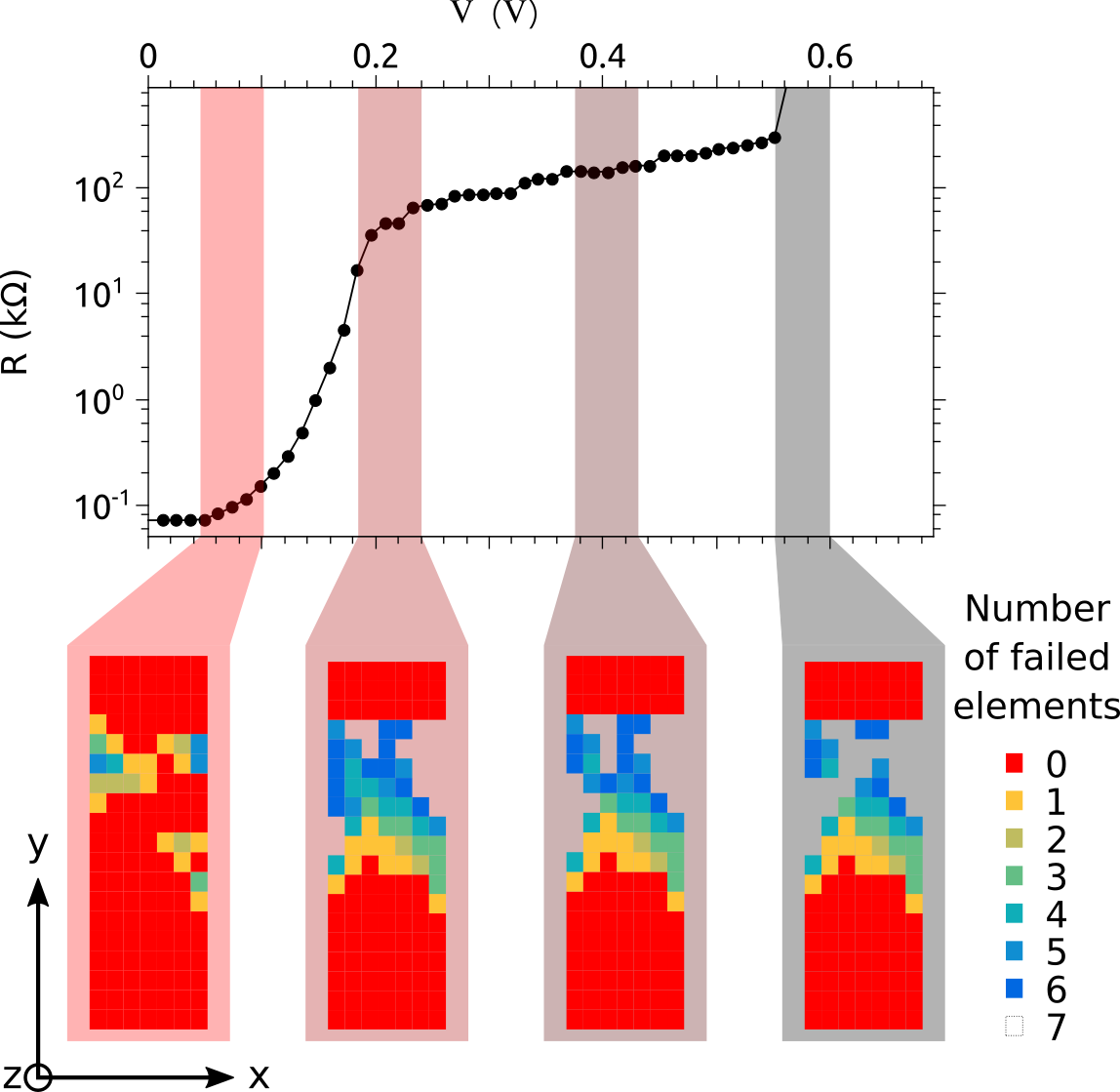}
\caption{Replicating the experiment on the effects of electromigration
on gold nanojunctions of Ref. \onlinecite{wu2007feedback} using our model for $R(I)$.
The values of the parameters have been chosen to match that experiment
and then used in our simulation of a ng-film.}
\label{fig:em}
\end{figure}

With this we are able to qualitatively reproduce the curves 
in \cite{wu2007feedback} in which a polycrystalline Au nanowire $100$ nm wide
and $200$ nm long ($7$ and $14$ ERN cells, respectively) is first activated
under $0.15$ V to finally reach the electrical failure at $0.4$ V. 
The curve obtained with our model is shown in Figure \ref{fig:em}: 
at low voltages, $V<0.05$ V, the sample displays a low resistance value ($R<$100 $\Omega$). At $V=0.1$ V some electromigration events occur 
(yellowish and bluish elements in the first panel in Figure \ref{fig:em}) 
and, as a consequence, the sample becomes more resistive 
but still in the low resistance range ($<1$ k$\Omega$). 
At $V=0.2$ V EM events have brought the sample to a high resistive state
($10-100$ k$\Omega$) and some gaps can be identified as not colored grid
elements in second to forth panel. 
Further increasing of the voltage produce the failure of an important portion
of the grid cells ending up with a short-circuited gold sample. 
Data shown in Figure \ref{fig:em} are obtained with $I_{a}$ and $I_{d}$
uniformly distributed between 19 $\mu$A to 23 $\mu$A and 32 $\mu$A 
and 38 $\mu$A, respectively. The values for $I_a$ at $t=0$ are chosen 
so that average current flowing through the sample just before 
the activation is in the order of magnitude of the measured one, 
i.e. $7$ mA. 
The upper limit for the $I_d$ range is set to be twice 
the lower limit for the $I_a$ one.

In a ng-film, where junctions are interfaces between adjacent clusters,
we hypothesize that a similar phenomenon happens.
More specifically, when the voltage is sufficiently high,
electromigration leads to a reduction of the cluster-cluster interface,
up until the two clusters actually separate.
We can therefore use the model we have outlined above also for ng-films.
The only additional feature we introduce is that, within each cell,
we allow junctions to shrink/break in each direction independently,
to account for the higher degree of inhomogeneity characterizing small length scales of a ng-film.
Indeed, in our picture, if in a cell two clusters get separated
along, say, the main direction of the current,
there might still be within the same cell a path connecting other clusters
in some other direction. 
It follows that the resistance of the cell along the direction of the bias, say $x$, is infinite,
$R_{x}=\infty$, but $R_{y}$ and $R_{z}$ are not, until, at least, 
the transversal current is not as strong.

Finally, we would like to emphasise that the modeling here proposed
neglects one important piece of information: the timescale over which
those effects are supposed to emerge (notice that Fig. \ref{fig:em}
has no indication of time). As pointed out in literature \cite{wu2007feedback},
the damage of even failure of a nanojunction may occur over a period of hours.
On the one hand, we do not have enough information to appropriately characterise the phenomenon in time;
on the other hand, such a timescale is quite far from that of the phenomena we are interested in (of the order of seconds, at most).
For now we make a strong assumption, to  take electromigration effects as instantaneous.
Later one, we shall come back on it and and offer the reader, on the basis of some concrete results,
an explanation on why such an assumption is in fact good enough for our purposes.

\section{Results}
\label{sec:results}

\subsection{Model Parameters}
\label{sec:modelparameters}

The model just described can be computationally demanding when the evolution 
of a ng-film is simulated in the experimental conditions.
Nevertheless, interesting insights can be gained already by considering smaller length and timescales.
By running several simulations of hundreds timesteps on samples of a few thousands of cells
under different conditions we were able to see some distinctive effects that can help understand 
the unusual behavior of ng-films.

Here we present a specific simulation that well captures some interesting, general trends observed.
We consider a uniform mesh of $80\times 80 \times 4$ cubic cells of side $d=15$ nm.
A specific ng-film sample is identified by the set of parameters that characterize the cells.
In our case, at $t=0$ each cell is characterized by:
an initial resistance tensor $R_{a;i,j,k}^{(0)}$ ($a=x,y,z$, $i=1,...,80$, $j=1,...,80$, and $k=1,...,4$)
randomly picked between $0$ and $250\;\Omega$;
a temperature threshold randomly picked between $300$ and $340$ K, and a resistance variation for the thermally-induced structural changes following \ref{eq:RofT} with $\Delta$ between $0$ and $0.2$;
two current thresholds randomly picked between $0.9 \mu A - 1.08 \mu A$ and $2.0 \mu A - 2.4 \mu A$ a shrinking factor $\beta=5$.
The value assumed for the remaining parameters appearing in the equations are $\rho c=0.016 kg^2*K^{-1}*m^{-1}*s^{-2}$ and $L=2.44\times 10^{-8}V^2 K^{-2}$.
To ensure convergence of the finite element scheme
a time step of $\tau=1$ ps has been adopted.
We allow heat to escape from the sample through all the boudaries of the sample by setting the temperature for those equal to their adjacent elements, i.e. $\frac{\partial T}{\partial a}\vert_{boundary}=0$ ($a=x,y,z$) and the temperature for all the elements is initially set to $T=300$ K. Finally, as bias we consider an applied voltage along the $x$ direction
that is ramped up from 0 to $0.014$ V, which produces an electric field close to the experimental one, during the first $30$ timesteps 
and kept constant after that. 
A simulation as such can require up to a few hours on a dual-core laptop.

Some of the above parameters differ from our own estimates presented in the above sections.
They have been tuned in order to explore an interesting region of the parameter space
within a reasonable simulation time (for instance, temperature thresholds are much lower
of those mentioned in Section \ref{sec:temp}). 
As we shall argue, this prevents us to interpret our results in an absolute, quantitative manner,
but not from getting some relative, qualitative insights.

\subsection{Sample evolution}

\begin{figure*}
\includegraphics[width=0.96\textwidth]{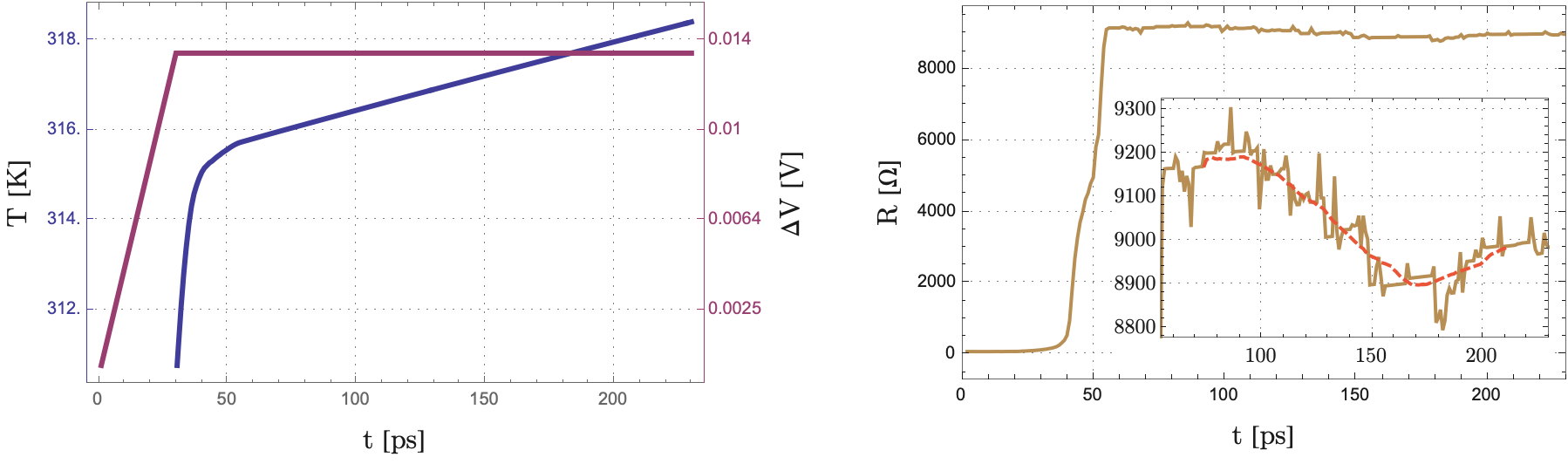}
\caption{Temperature (blue line) and resistance (gold line) of a simulated ng-film.
In the left panel, electrodes voltage difference is also plotted (magenta line)
for reference. In the right panel, an inset plot shows a zoom-in
of the resistance after the activation phase. 
Dashed line in the inset represents the resistance value averaged over 
20 previous and 20 following timesteps.}
\label{fig:sampleRandT}
\end{figure*}

In the left panel of Fig. \ref{fig:sampleRandT} the evolution of the sample resistance over time is reported.
We distinctly recognize two phases, connected with the external voltage.
When the voltage is raised, the resistance rises, too, ultimately reaching a value that is
orders of magnitudes higher than the initial one.
Once the voltage is constant, the resistance stabilizes,
although, on a smaller scale, some variations are still present.

\begin{figure*}
\centering
\includegraphics[width=0.9\textwidth]{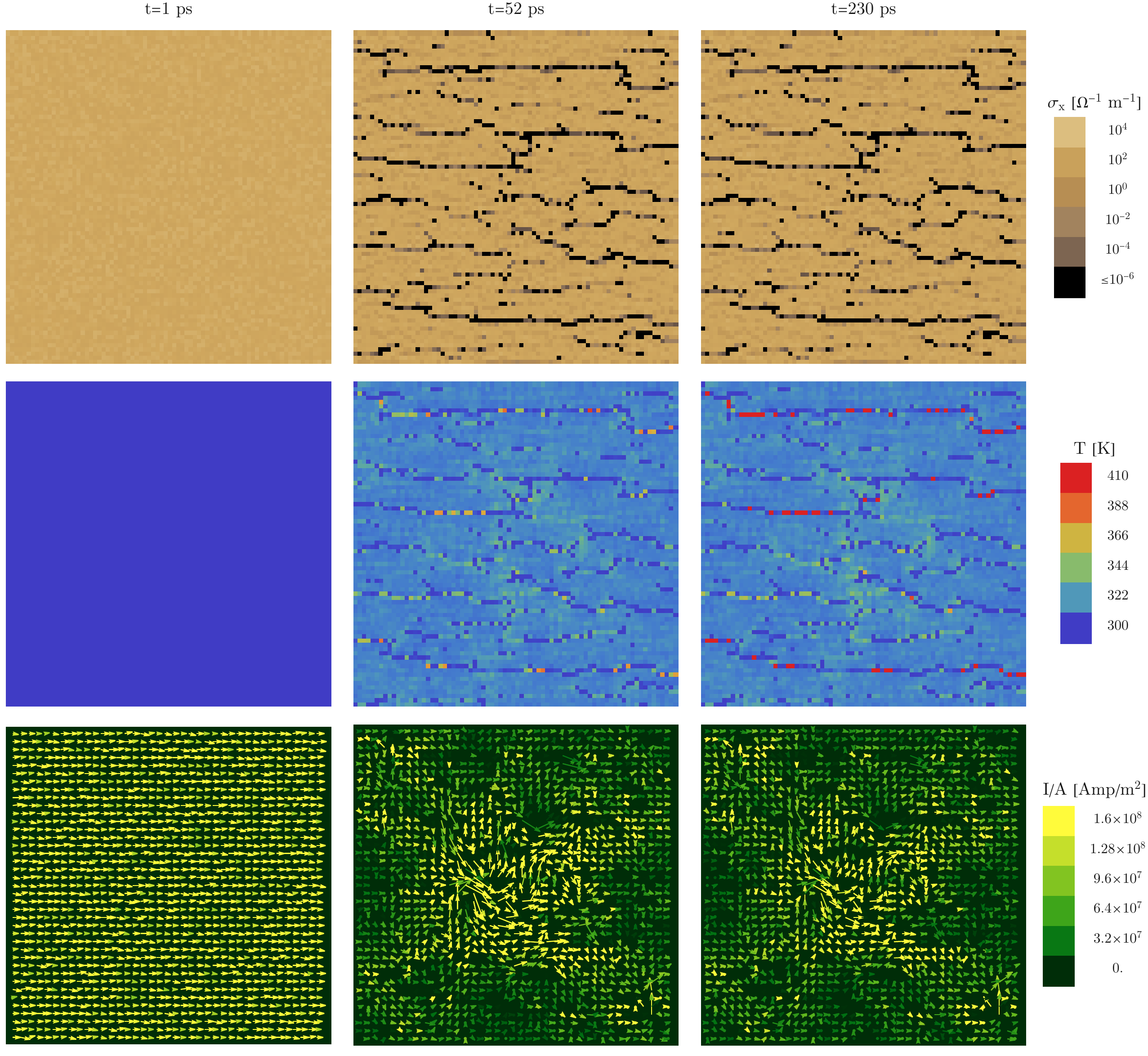}
\caption{Conductivity along the bias direction (first row), 
temperature (second row),
and current field (third row) inside the simulated ng-film 
at three different time steps.
The first time step correspond to the beginning of the simulation
when the sample is at room temperature, its conductivity is high
and the current flows rather uniformly from one electrode to the other;
the second plotted step corresponds to the end of the deterioration phase induced by electromigration effects;
the last plotted step is the last simulated one, which shows that
changes occurring the second phase are much less visible.
Electrodes are placed at the left and right sides of a square;
the section is taken at half the ng-film thickness;
the arrows in the third row represent the project of the current
over that section; to better visualize the current field,
both the length and the color of the arrows represent its intensity.
}
\label{fig:maps}
\end{figure*}

As we analyze the change of resistance of each single cell, 
we recognize that the initial phase is dominated by electromigration effects.
They lead to an extensive deterioration of the sample, as shown in the top row of Fig. \ref{fig:maps}.
Nearly a third of the cells experienced a shrinkage of an interface along the bias direction, 
and in most cases ($\sim 80\%$) the shrinkage turned into a broken connection. 
As an effect of the current trying to find its way through such an inhomogeneous medium,
also interfaces along the transverse direction were highly affected by electromigration. 
Nearly $12\%$ of cells experienced interface shrinkage along one
or the other transverse direction and $65\%$ of them completely lost the connection.
Nevertheless, only $1.5\%$ of cells were left completely insulating, 
i.e. with no current flowing in any direction.
The voltage was chosen to be not high enough to actually break the sample,
so at some point the system stabilizes and electromigration 
effects become negligible.

After that first phase, thermal effects dominates. 
On small time scales they lead to some noisy variations of the film resistance. 
But on a larger time scale, they seem to coordinate to form larger and more stable resistance variations,
that one can appreciate in the inset plot of the left panel of Fig \ref{fig:sampleRandT}.

From the right panel of Fig. \ref{fig:sampleRandT} we can see the evolution of the sample temperature. 
As one would expect from the boundary conditions, it always rises.
Also, consistently with the two regimes for the sample resistance,
such a rising happens at two different rates: a high one in the beginning,
when the sample resistance is relatively low, and a low one,
when the resistance is high.
Such a behavior is direct consequence of the fact that the increase in temperature
is inversely proportional to $R^3$, a relation one can obtain by combining the power
dissipated by Joule effect, $P=I^2/R$, and Ohm's first law, $I=V/R$.
In the local temperature map plotted in the second row of Fig. \ref{fig:maps}, 
we also see that spots with higher resistance tend to become hotter. 
This is due to the fact that, as the spot electrical conductance lowers,
so does its thermal one (the two are linked by Eqn. \eqref{eq:wiedemannfranz});
and because of the slow heat dissipation rate,
the spot overheats.

We notice that the rise in temperature in the second phase is, rather interestingly, 
somehow at odds with the trend of the resistance.
In a simplified picture, one would expect that, 
as the sample temperature goes to higher values,
cell resistances, which are subject to Eq. \ref{eq:RofT}, 
would, on average, tend to rise as well;
and, in turn, the sample resistance would rise, too. But this is not what happens.
Although cell resistances do increase on average,
the sample resistance, which is calculated using the circuit law,
presents a rather different trend.

In the last row of Fig. \ref{fig:maps} a comparison between the current before and after 
the sample deterioration is presented.
Before the deterioration occurs, 
the current basically flows uniformly from one electrode to the other.
But once the deterioration is over, clear preferred paths emerge.
Although only few cells are completely insulating ($\sigma_x=\sigma_y=\sigma_z=0$),
the fact that a third of the cells does not let the current flow along the bias direction
(only $\sigma_x=0$) shapes the current in a very specific way.

\subsection{Connection with experiments}

To provide a physical interpretation of the simulation 
and compare it with experiments we need to remind the reader
that the time scales of the simulations are not realistic,
in two ways.
First, as discussed in Sec. \ref{sec:curr},
the way $R(I)$ has been modeled on electromigration experiments
completely ignores the typical time scales of the phenomenon,
for which too little information is given to know.
It follows that the time scale of the first phase is somehow artificial,
for it should almost certainly happen on macroscopic scales.
Second, although we got some realistic estimates for the parameters of the function $R(T)$,
we deliberately altered them in order to probe interesting regions 
of the parameter space within a reasonable simulation time.
Less dramatically than for the first phase,
in the second phase the time scale, although not quite realistic,
might well be, under certain conditions of external voltage and sample temperature, 
not that far from a realistic one. 

Bearing that in mind, we see from the above analysis that,
even though the two effects are treated, as we did, on the same timescale,
they are distinct and consequential, 
with thermal effects being relevant only after
electromigration has deteriorated the sample.
In fact, this is representative of a more general trend
we observed in our simulations.
If electromigration effects are artificially switched off, 
thermally triggered local resistivity variations have little effects on the overall sample resistance.
In other words, local resistance variations triggered by the temperature can have 
a macroscopic relevance only if the sample has previously been deteriorated 
by electromigration effects. It seems therefore reasonable to draw a direct comparison
between the deterioration phase that we observe in the simulations
and the sample ``activation'' observed in the experiments:
in both cases the sample resistance rises by orders of magnitudes,
it represent a necessary phase to go through for the emergence of resistive switching events at the macroscopic scale,
and it is permanent.

As mentioned, in the second phase, when electromigration plays no longer a role,
thermally activated local resistance variations can coordinate 
to lead to macroscopically appreciable ones. 
Those variations are actually also in a quantitative agreement with experiments:
in our simulation the sample goes from 
an average resistance of $ 9200 \; \Omega$ to $8900 \; \Omega$,
which is indeed strikingly close to the resistance values explored by one of the sample
discussed in Ref. \onlinecite{mirigliano2019}, ranging from $9900\;\Omega$
to $9300\;\Omega$.\footnote{%
We recall the reader that, even though our sample is much smaller than that of the experiments,
the proportions are kept in such a way the resistance is the same.
}
Time scale is indicative of a microscopic phenomenon, 
meaning that what here appears as a smooth variation,
it would appears as a rapid transition on macroscopic scales,
in agreement with the ``sudden'' jumps reported in experiments.

A piece of explanation of the reason why resistive switching events emerge 
on deteriorated samples comes from the analysis of the current field.
As mentioned, on the scales considered, the deterioration leads to the emergence of a 
pattern of current, which does not flow uniformly from one electrode to the others 
but via some preferred paths.
And, just like in the submonolayer regime when only a few percolation paths are available,
also here it is true that even slight changes on a path can have macroscopic effects.
If, for instance, the switching of just a few cells to a high resistive state
makes a path no longer favourable,
such an event can lead to a sensible resistance variation of the entire sample
if the current has only limited options to run from one side to the other.

Under different experimental conditions, 
jumps in real samples present different features (cf panels (a) and (c)
of Fig. 4 in Ref. \onlinecite{mirigliano2019}) 
that may be sign of different microscopic origins. 
From current available information it is hard, if not impossible, 
to say whether the jumps we observe in our simulations are caused by the same mechanisms. 
Only a synergetic effort of theory and experiment in characterizing features 
such as recurrent values, plateauing duration, etc. in connection with the experimental setup 
and ng-film specs will enable a more precise characterisation 
of the relevant microscopical mechanism(s).


\section{Conclusions and Outlook}
\label{sec:conc}

In this work we presented a way to model the electric response
of large-scale complex nanostructures with local structural variations.
The problem has been tackled at different scales with different tools.
At the macroscopic level, a set of classical equations (the so-called thermistor problem)
have been used to get the evolution in time of the
system temperature and resistance
in presence of local variations of conductivity.
We then considered molecular dynamics simulations
to characterize such variations when arising from 
thermally activated microscopic structural rearrangements.
This was made possible by a tool, presented already in Ref. \onlinecite{lopez2021},
that is based on an \textit{ab initio} characterization 
of electronic transport at the atomic scale.
Furthermore, electromigration effects, that were modeled on relevant experiments,
were also included.
Altogether, the model allowed us to simulate the emergence of resistive switching
events in nanogranular gold films.

We identified electromigration as responsible of 
an extensive deterioration of the sample
occurring in experiments during the so called activation phase.
Such a deterioration forces the current to flow only on a few preferred paths
and amplifies the effect of local variations of resistance
which can indeed have macroscopic relevance.
Furthermore, in the measurement of the sample resistance over time,
local resistance variations appear to coordinate and lead to large variations that,
on macroscopic timescales, can be identified as resistive switching event.

Such encouraging results certainly motivate to extend our work
in several directions:
running larger scale (both of time and length) simulations in search of
other key features of resistive switching in ng-films;
exploring larger regions of the parameter space
(also covering possible-but-yet-to-explore experimental conditions);
refine the models for $\tensor{\sigma}(T,\vec{I})$
using more extensive and systematic MD simulations and other theoretical tools; 
and, last but not least, apply the same methodology to other systems
(submonolayer ng-films, more complex, or composite nanostructures, etc.).

\section*{Appendix}

\subsection*{Numerical resolution of the thermistor problem}

Once the map $T,\vec{I}\to \tensor{\sigma}$ has been characterize,
the set of equations \eqref{eq:fourier}, \eqref{eq:kirchhoff}, \eqref{eq:wiedemannfranz}
supplied with appropriate initial and boundary conditions
represent a well-defined problem that can be solved numerically.
We use a regular mesh of cubic cells, $\Delta x \times \Delta y \times \Delta z = d^3$. 
This implies that the current can only flow in three directions through the mesh element 
(along $x$, $y$, or $z$), i.e. $\sigma_{ab}\to \sigma_a$ $(a=x,y,z)$.
With that simplification the heat equation reduces to
\be
\rho c \frac{\partial T}{\partial t}=\partial_a \kappa_a \partial_a T+{I_a}^2/\sigma_a
\ee
where Einstein summation notation over $a=x,y,z$ is understood.

As pointed out in Ref. \onlinecite{praprotnik2004}, one has to slightly change the algorithm 
for solving the heat equation in dealing with an inhomogeneous thermal conductivity. 
Using the anisotropic generalization of their Eqn.'s 30 and 31,
we have that an explicit (in time) finite element version of the above equation is
\begin{widetext}
\begin{multline}
\rho_{i,j,k}c_{p_{i,j,k}}\frac{T^{(n+1)}_{i,j,k}-T^{(n)}_{i,j,k}}{\Delta t}=
2\left(
\kappa^{(n)}_{x;i,j,k}\frac{T^{(n)}_{x;i+1/2,j,k}-2T^{(n)}_{i,j,k}+T^{(n)}_{x;i-1/2,j,k}}{\Delta x^2}+\right.\\
+\left.\kappa^{(n)}_{y;i,j,k}\frac{T^{(n)}_{y;i,j+1/2,k}-2T^{(n)}_{i,j,k}+T^{(n)}_{y;i,j-1/2,k}}{\Delta y^2}+
\kappa^{(n)}_{z;i,j,k}\frac{T^{(n)}_{z;i,j,k+1/2}-2T^{(n)}_{i,j,k}+T^{(n)}_{z;i,j,k-1/2}}{\Delta z^2}
\right)+\\
+{I_{x;i,j,k}^{(n)}}^2/\sigma^{(n)}_{x;i,j,k}+{I_{y;i,j,k}^{(n)}}^2/\sigma^{(n)}_{y;i,j,k}+{I_{z;i,j,k}^{(n)}}^2/\sigma^{(n)}_{z;i,j,k}
\end{multline}
\end{widetext}
with
\be\begin{array}{l}
T^{(n)}_{x;i\pm 2,j,k}=\frac{
\kappa^{(n)}_{x;i,j,k}T^{(n)}_{i,j,k}+\kappa^{(n)}_{x;i\pm 1,j,k}T^{(n)}_{i\pm 1,j,k}
}{
\kappa^{(n)}_{x;i,j,k}+\kappa^{(n)}_{x;i\pm 1,j,k}
}\\
T^{(n)}_{y;i,j\pm 2,k}=\frac{
\kappa^{(n)}_{y;i,j,k}T^{(n)}_{i,j,k}+\kappa^{(n)}_{y;i,j\pm 1,k}T^{(n)}_{i,j\pm 1,k}
}{
\kappa^{(n)}_{y;i,j,k}+\kappa^{(n)}_{y;i,j\pm 1,k}
}\\
T^{(n)}_{z;i,j,k\pm 2}=\frac{
\kappa^{(n)}_{z;i,j,k}T^{(n)}_{i,j,k}+\kappa^{(n)}_{z;i,j,k\pm 1}T^{(n)}_{i,j,k\pm 1}
}{
\kappa^{(n)}_{z;i,j,k}+\kappa^{(n)}_{z;i,j,k\pm 1}.
}
\end{array}\ee
where $n$ is the timestep index and $i,j,k$ are the cell indices along the $x,y,z$ directions, respectively.
With our cubic mesh, the electrical conductivity reduces to
\be
\sigma^{(n)}_{a;i,j,k}=1/(d\;R^{(n)}_{a;i,j,k})
\ee
while the thermal conductivity, for which we use Eq. \eqref{eq:wiedemannfranz},
is 
\be
\kappa^{(n)}_{a;i,j,k} = L\; T^{(n)}_{i,j,k} /(d\;R^{(n)}_{a;i,j,k}).
\ee
$R$ being the cell resistance.

The initial value is imposed by assigning a value to the entire temperature field.
In experimental conditions, one would start from a sample at environment temperature: 
$T^{(0)}_{i,j,k}=T_{environment}$ for all $i,j,k$.
When Dirichlet boundary conditions are chosen, one should impose 
\be \begin{array}{l}
T^{(n)}_{i_{min/max},j,k}=T_{environment},\\
T^{(n)}_{i,j_{min/max},k}=T_{environment},\\
T^{(n)}_{i,j,k_{min/max}}=T_{environment}
\end{array}\ee
for all free indices at each time step.
For Neumann boundary conditions, one imposes
\be\begin{array}{l}
T^{(n)}_{i_{min},j,k}=T^{(n)}_{i_{min}+1,j,k}\\
T^{(n)}_{i_{max},j,k}=T^{(n)}_{i_{max}-1,j,k}\\
T^{(n)}_{i,j_{min},k}=T^{(n)}_{i,j_{min}+1,k}\\
T^{(n)}_{i,j_{max},k}=T^{(n)}_{i,j_{max}-1,k}\\
T^{(n)}_{i,j,k_{min}}=T^{(n)}_{i,j,k_{min}+1}\\
T^{(n)}_{i,j,k_{max}}=T^{(n)}_{i,j,k_{max}-1}
\end{array} \ee
for all free indices at each time step.
Mixed boundary conditions, or even more complicated ones,
can be used to model the presence of the electrodes,
the substrate and the heat dissipation of the upper surface.

About the circuit equation. Eqn. \eqref{eq:kirchhoff},
one can use the scheme we presented in Ref. \onlinecite{lopez2021}, 
in which we actually use the integrated version of the stated circuit equation 
(sum of cell incoming currents =0) which can be obtained using Gauss's Divergence Theorem 
to the continuity equation assuming no accumulation/depletion of charges ($\partial_t q=0$). 
See that paper for further information about the numerical resolution.
In that framework, the current is rewritten in terms of the difference of potentials 
at the nodes and the cell resistances, 
$\vec{I}=-\overleftrightarrow{\sigma}\cdot \vec{\nabla} v$
which becomes
$I_{x;i,j,k}=-(v_{i+1,j,k}-v_{i,j,k})/(d^2 R_{a;i,j,k})$.
It should be mentioned that there is an underlying assumption of staticity of the current, 
even when the jumps occur, which is here deemed as reasonable for the electronic dynamics 
is characterized by scales time quite far from those here considered.

\subsection*{Molecular dynamics simulation and resistance estimates
of carved samples}

\begin{figure}
\centering
\includegraphics[width=0.48\textwidth]{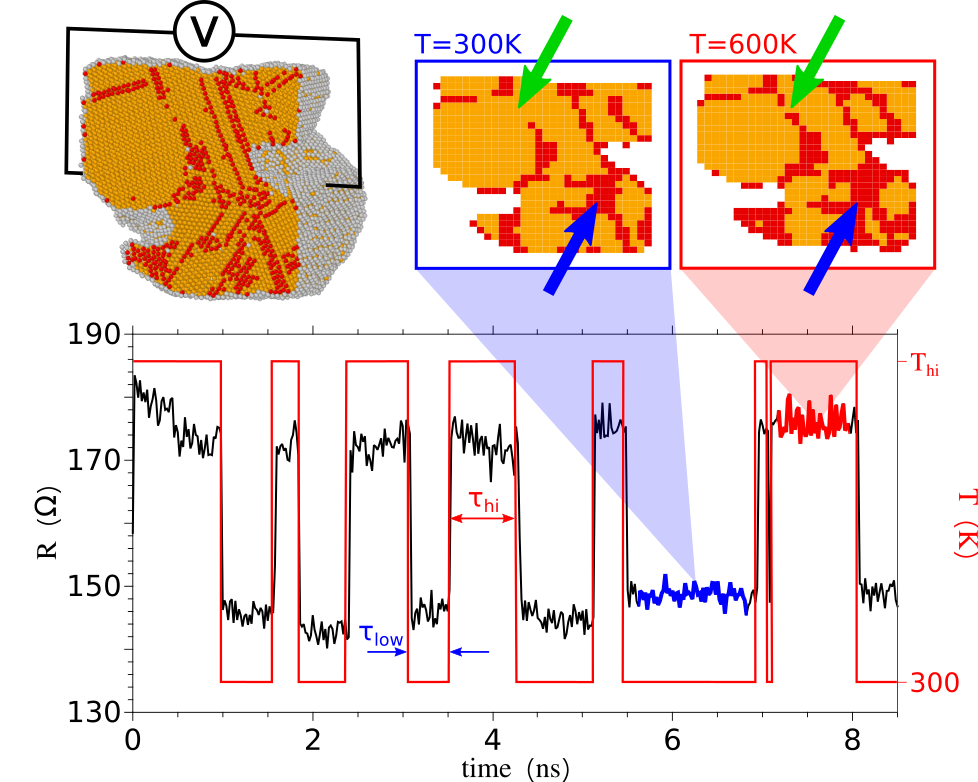}
\caption{Time evolution of the computed $R$ value for a single carved junction.
Upper panels. Left: Section view of a single junction. A Polyhedral Template Matching Analysis allows to localize with atomic resolution the distribution of the defects within the junction: orange atoms are found in fcc sites while red ones cannot be classified as such. Middle: Averaged PTM at $T=300K$. Right: Averaged PTM at $T=600K$. Lower panel: $R$ (black) and temperature (red) time series showing the effect of temperature modulation on the computed $R$ and the reversibility of such dynamics. The AR-ERN is applied every $20$ ps while the time step for the MD simulation is $1$ fs.}
\label{fig:fig4}
\end{figure}

MD simulations of the three systems presented in Section \ref{sec:temp} have been performed using the LAMMPS code \cite{plimpton1993fast}, integrating the equations of motion by the velocity-Verlet algorithm. The Au-Au interactions were sampled using the embedded-atom potential of Ref. \onlinecite{foiles1986embedded}. By means of a Nose-Hoover thermostat (relaxation time  equal  to  100  fs) we are able to dynamically modulate the temperature, $T$, of each of these gold systems (or ``junctions'') while keeping track of the fcc and non-fcc regions evolution. We opt for a two-level switching scheme, meaning that we will modulate the junction temperature between two fixed values, $T_{room}=300$ K and $T_{hi}$.
We expect the micro-structural changes to be more pronounced for higher temperatures. Unfortunately, we have no hint on the local or averaged temperature of the experimental samples in \cite{mirigliano2019}. However, other studies have treated the relation between EM and temperature in gold nano-junctions: maximum temperatures in the range of $450$K \cite{stahlmecke2007resistance,wu2007feedback} and $850$ K \cite{karim2009diameter} have been reported at moderate current densities, i.e. $1.5 \cdot 10^{12}$ and $0.3 \cdot 10^{12}$ A/m$^2$, respectively. Consistently, we set $T_{hi}$ to $430$ K and $600$ K, the later representing a maximum temperature well below the melting point of the simulated Au junctions ($\sim900$ K) and below the reported maximum temperatures. Every junction is initially left to evolve for $1.0$ ns at $T=300$ K. Next, the thermostat temperature is switched from $T=300$ K to $T_{hi}$ and the junction is left to evolve again at its new temperature for a period $\tau_{hi}$. Finally, the thermostat is switched back to $T=300$ K, this time for a period $\tau_{room}$. This operation is repeated varying randomly the length of $\tau_{hi}$ and $\tau_{low}$, as shown in Fig. \ref{fig:fig4}. We use such a scheme rather than a periodic switching scheme to avoid inducing any kind of resonances or other spurious effects. Atoms belonging to the edges of each junction are kept in their initial positions during the simulation at finite temperature. This is done in order to prevent the whole structure to completely relax to a single fcc domain and mimic the presence of the macroscopic granular film.

Every carved sample contains two kinds of atomic arrangements: cubic (fcc) and non-cubic (non-fcc) gold, colored in orange and red respectively in Fig. \ref{fig:fig4}: a Polyhedral Template Matching (PTM) Analysis performed with Ovito software \cite{stukowski2009visualization} allows to distinguish between those Au atoms sitting in fcc sites (forming fcc regions) and those which are not (forming non-fcc regions). In Fig. \ref{fig:fig4} we show a section of a carved sample where atoms are colored according to the PTM analysis. We observe non-fcc regions (red colored atoms) separating different fcc regions (orange colored atoms) within a single junction. As explained in \cite{lopez2021modeling}, fcc domains are less resistive than non-fcc ones, thus, less defected junctions tend to conduct better. Moreover, the specific distribution of the fcc and non-fcc regions determines the resulting $R$ value for the junction.

The color maps displayed in the right-upper panels of Fig. \ref{fig:fig4} show the fcc and non-fcc distribution averaged over the ERN cells at two different time steps corresponding to $T=300$ K and $T=600$ K. We observe that a temperature increase produces two effects: (i) a redistribution of the structural defects and (ii) a broadening of non-fcc regions. In Fig. \ref{fig:fig4} these features are highlighted with green and blue arrows, respectively. As a consequence the ratio between fcc and non-fcc cells increases and these variations modulate the electrical response of the junction itself.

The total resistance of each junction is calculated 
by means of the AR-ERN of Ref.\onlinecite{lopez2021modeling}, 
every $2 \cdot 10^4$ MD time-steps. As shown in Fig. \ref{fig:fig4}, the total resistance is effectively modulated with the sample temperature in a reversible manner. 
The same MD protocol has been used for $T_{hi}=430$ K and $600$ K and three different gold junctions, all displaying similar behavior as shown in Fig. \ref{fig:cycle}: the total junction resistance is modulated by the sample temperature. The observed jumps in the $R$ value, $\Delta R$, reach maximum values in the order of a $25\%$ of its initial value. Importantly, not all junctions perform equally: the junction corresponding to the upper panel shows very little $R$ variations for the $T_{hi}=430$ K case ($\Delta R<5\%$), while the modulation is much more pronounced for the $T_{hi}=600$ K case ($\Delta R \sim 25\%$). The same feature is observed in the middle panel. Instead the junction corresponding to the lower panel responds similar for the two thermal modulation conditions providing a maximum $\Delta R$ equal to $7\%$ and $10\%$ for $T_{hi}=430$ K and $T_{hi}=600$ K, respectively.

\begin{acknowledgments}
This work was fully funded by Fondazione CON IL SUD (Grant No: 2018-PDR-01004).
\end{acknowledgments}
\bibliography{bibliography_new.bib}
\end{document}